\documentclass[prb,twocolumn,showpacs,amssymb]{revtex4}

\usepackage{graphicx}
\usepackage{dcolumn}
\usepackage{bm}

\begin{document}


\title{Thomas-Fermi scaling in the energy spectra of atomic ions}
\author{Robert Carcass\'es$^a$ and Augusto Gonz\'alez$^b$}
\affiliation{$^a$ Universidad de las Ciencias Inform\'aticas, Cuba\\
$^b$ Instituto de Cibern\'etica, Matem\'atica
 y F\'isica, Calle E 309, Vedado, Ciudad Habana, Cuba}

\begin{abstract}
The energy spectra of atomic ions are re-examined from the point of view
of Thomas-Fermi scaling relations. For the first ionization potential,
which  sets the energy scale for the true discrete spectrum, 
Thomas-Fermi theory predicts the following relation: 
$E_{ioniz}=Z^2 N^{-2/3} g(N/Z)$, where $Z$ is the nuclear charge, $N$ 
is the number of electrons, and $g$ is a function of $N/Z$. This 
relation does not hold for neutral atoms, but works extremely well in 
the cationic domain, $Z>N$. We provide an analytic expression for $g$, 
with two adjustable parameters, which fits the available experimental 
data for more than 380 ions. In addition, we show that a rough fit to 
the integrated density of states with a single exponential: 
$N_{states}=\exp (\Delta E/\Theta)$, where $\Delta E$ is the excitation 
energy, leads to a parameter, $\Theta$, exhibiting a universal scaling 
{\it a la} Thomas-Fermi: $\Theta=Z^2 N^{-4/3} h(N/Z)$ , where $h$ is 
approximately linear near $N/Z=1$.
\end{abstract}


\maketitle

Energy levels of atoms and ions have been measured since the earliest
times of Quantum Mechanics. At present, there is a huge amount of very
precise measurements, which can be found, for example, in the critical 
compilation by the NIST \cite{NIST}. In these big volumes of data,
sometimes universal relations remain hidden. In order to find them, you
need a simple theory in which universality or scaling naturally appears.

Thomas-Fermi theory \cite{TF} was the first 
mean-field theory of atoms, the predecessor of more elaborated 
approaches, such as Density Functional Theory \cite{dft}. It is based on a
simple estimation of the kinetic energy, with the help of Pauli exclusion
principle, and the introduction of self-consistency with the Poisson
equation in order to take account of Coulomb interactions. Thomas-Fermi 
theory has proven to be a valuable tool for the 
qualitative and semi-quantitative understanding of real 
\cite{Kirzhnits,Lieb1,Spruch} and artificial \cite{Lieb2} atoms.

A very interesting aspect of Thomas-Fermi theory is the scaled form of
physical magnitudes, which are very often respected by the exact quantum 
mechanical magnitudes. This aspect has
not been completely exploited so far. Recently, we showed, for example,
that the density of low-lying excited states of artificial (quantum dot) 
atoms can be parametrized in a universal (scaled) way 
\cite{universality}, suggested by Thomas-Fermi theory. No similar 
parametrization for neutral atoms or ions is available yet.

In the present letter, we focus on the energy spectra of atomic ions, 
and show that, in some sense, the excitation spectrum is 
``universal'', that is, may be described by scaled functions. In order
to support the theoretical statements, we make extensive use of 
the NIST detailed compilation on atomic energy levels.

Let us start considering the first ionization potential of ions, which sets
the energy scale for the true discrete spectrum. In atomic units, where 
atomic ions are characterized by only two free parameters: the nuclear 
charge, $Z$, and the number of electrons, $N$, the ground-state energy 
in the Thomas-Fermi approximation may be shown to satisfy the following 
scaled relation\cite{Lieb1}:

\begin{equation}
E_{gs}(N,Z)=Z^2 N^{1/3} f(N/Z),
\label{eq1}
\end{equation}

\noindent
where $f$ is a universal function of the variable $N/Z$. The ionization
potential can be obtained from Eq. (\ref{eq1}). Indeed, 

\begin{eqnarray}
E_{ioniz}(N,Z)&=&E_{gs}(N-1,Z)-E_{gs}(N,Z)\nonumber\\
 &\approx& -\frac{\partial}{\partial N} E_{gs}(N,Z)\nonumber\\
 &\approx& Z^2 N^{-2/3} g(N/Z),
\label{eq2}
\end{eqnarray}

\noindent
where $g$ is also a universal function of its argument. It may be easily
verified for free electrons in a Coulomb field that, in the $N\to\infty,
~N/Z\to 0$ limit, $g(0)=(3/2)^{1/3}/3\approx 0.3816$. \cite{note}

\begin{figure}[t]
\begin{center}
\includegraphics[width=.95\linewidth,angle=0]{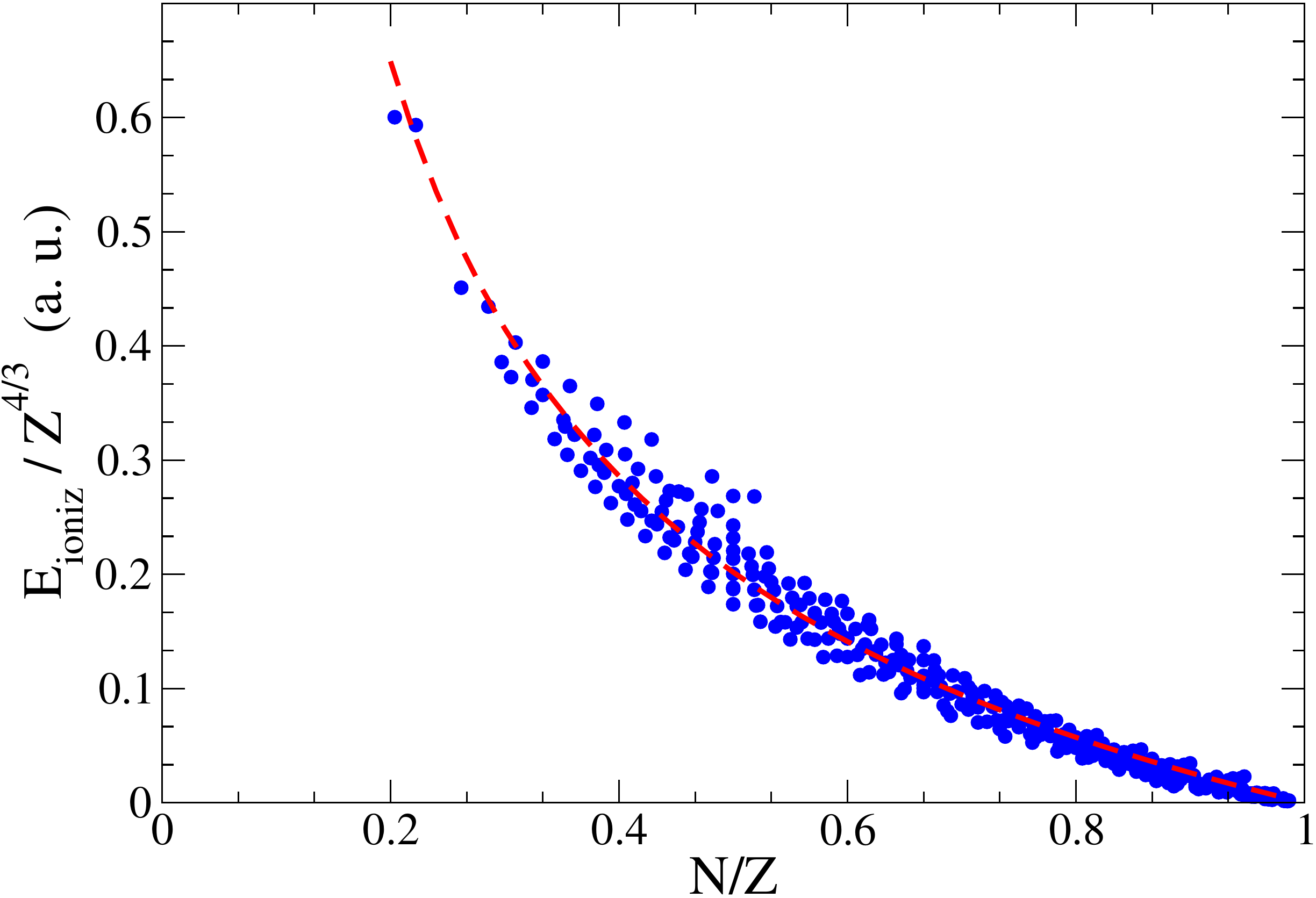}
\caption{\label{fig1} (Color online) Scaling in the ionization potential
of ions with more than ten electrons. The curve $(N/Z)^{-2/3} g$, where
$g$ is given by Eq. (\ref{eq3}), is also drawn (dashed line).}
\end{center}
\end{figure}

The law given by Eq. (\ref{eq2}) does not hold for neutral atoms. The 
ionization potential of neutral atoms,
apart from shell effects, has a very smooth dependence on $Z$, and is
close to that of Hydrogen, 0.5, in accordance to the fact that at long
distances every singly ionized atom looks like a proton.

\begin{figure}[t]
\begin{center}
\includegraphics[width=.95\linewidth,angle=0]{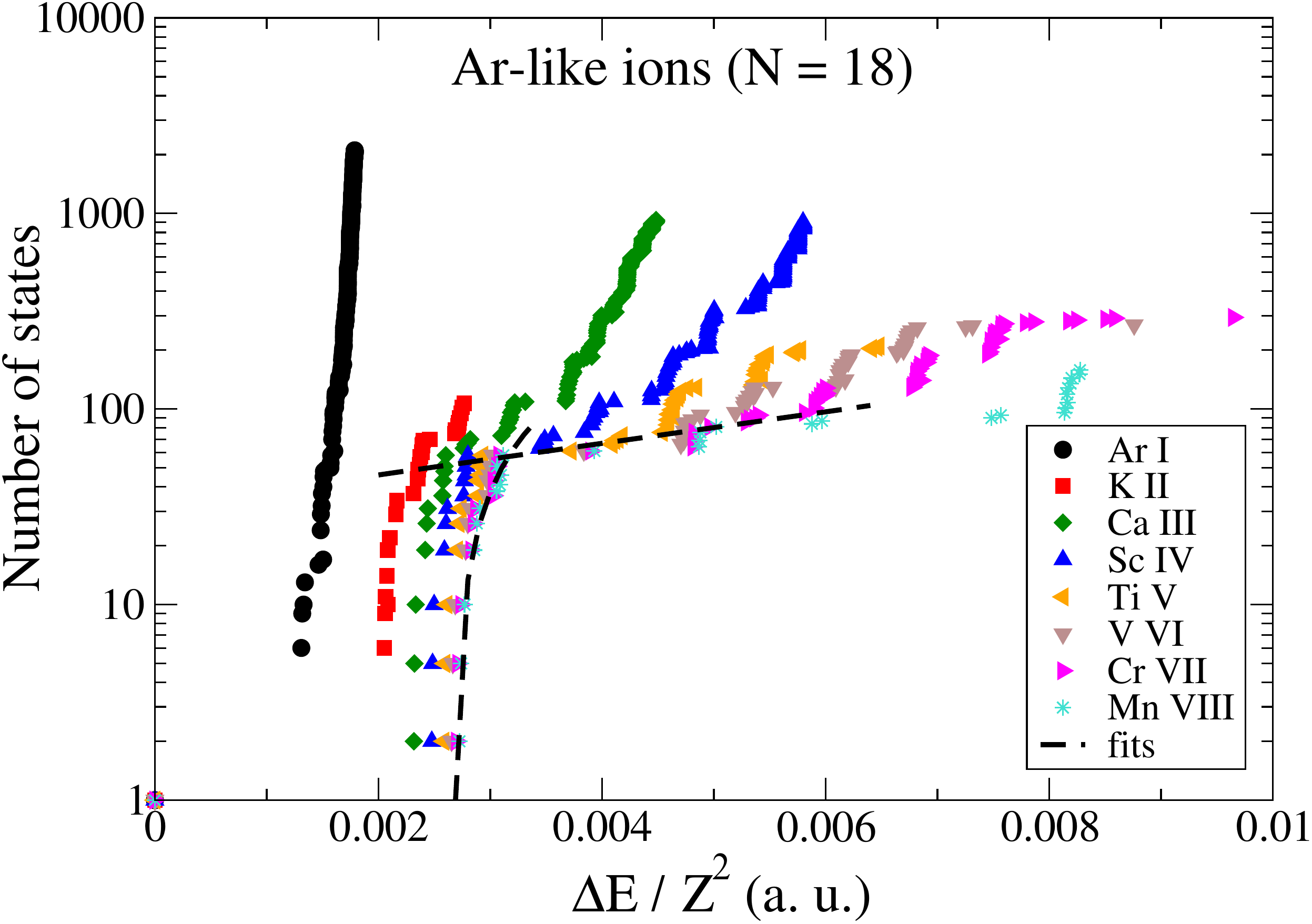}
\caption{\label{fig2} (Color online) The excitation spectra of Ar-like
ions showing the $Z^2$ dependence of intra-shell excitations, and the
exponential increase of $N_{states}$ with $\Delta E$.}
\end{center}
\end{figure}

However, the scaling law, Eq. (\ref{eq2}), is quickly reestablished in
the cationic domain, $Z>N$. We show in Fig. \ref{fig1} the ionization
potential of more than 380 atomic ions, taken from the NIST 
compilation \cite{NIST}. All of the systems with $N>10$ are considered, 
for which the scaled behaviour is apparent. The nuclear charge, $Z$,
spans the range from 12 (Mg) to 80 (Hg), although, as it can be seen
from the table of available levels \cite{levels}, most of the data 
correspond to the interval $12<Z<37$.

The function $g(x)$ should go to zero in the $x\to 1$ limit, in
accordance to the fact that, in the leading approximation, the
anionic instability border is located at $N/Z=1$.\cite{border} We 
fitted the experimental data with a two-point Pad\'e approximant 
\cite{pade} interpolating between the $x\to 0$ and $x\to 1$ limits:

\begin{equation} 
g(x)=\frac{(p_0+p_1 x)(1-x)}{1+q_1x},
\label{eq3}
\end{equation}

\noindent
where $p_0=g(0)$, and the fitting parameters are: $p_1=2.6555$,
$q_1=11.4399$. The fit gives a very good qualitative description of the
data. The three points at $N/Z\approx 0.5$, lying a little farther, 
correspond to the Xe XXVII, Xe XXVIII, and Xe XXIX ions, for which it 
seems that there is an overestimation in the reported values 
\cite{note2}. 

We stress that the fit given by the function $g$ captures the general 
behaviour of $E_{ioniz}$ as a function of $N$ and $Z$. Effects not
included in Thomas-Fermi theory, such as shell or sub-shell filling,
spin-orbit interactions, etc, could lead to relatively large (of the
order of 10\% or even larger) deviations from the universal 
behaviour, as seen in Fig. \ref{fig1}.

Next, we attempt a global universal description of the excitation 
spectrum.
Excitation energies, $\Delta E$, cover the range $(0,E_{ioniz})$.
In the spectrum, we shall distinguish between intra-shell and 
inter-shell excitations, which exhibit different behaviours.

In quality of example, we show in Fig. \ref{fig2} the Ar-like
series, from the Ar I atom ($N=Z=18$) to the Mn VIII ion ($N=18$,
$Z=25$). The $y$ axis represents the total (accumulative) number of
states for excitation energies below $\Delta E$. The degeneracy of
multiplets is explicitly taken into account. In the $x$ axis, the
excitation energy is scaled by $Z^2$.

\begin{figure}[t]
\begin{center}
\includegraphics[width=.95\linewidth,angle=0]{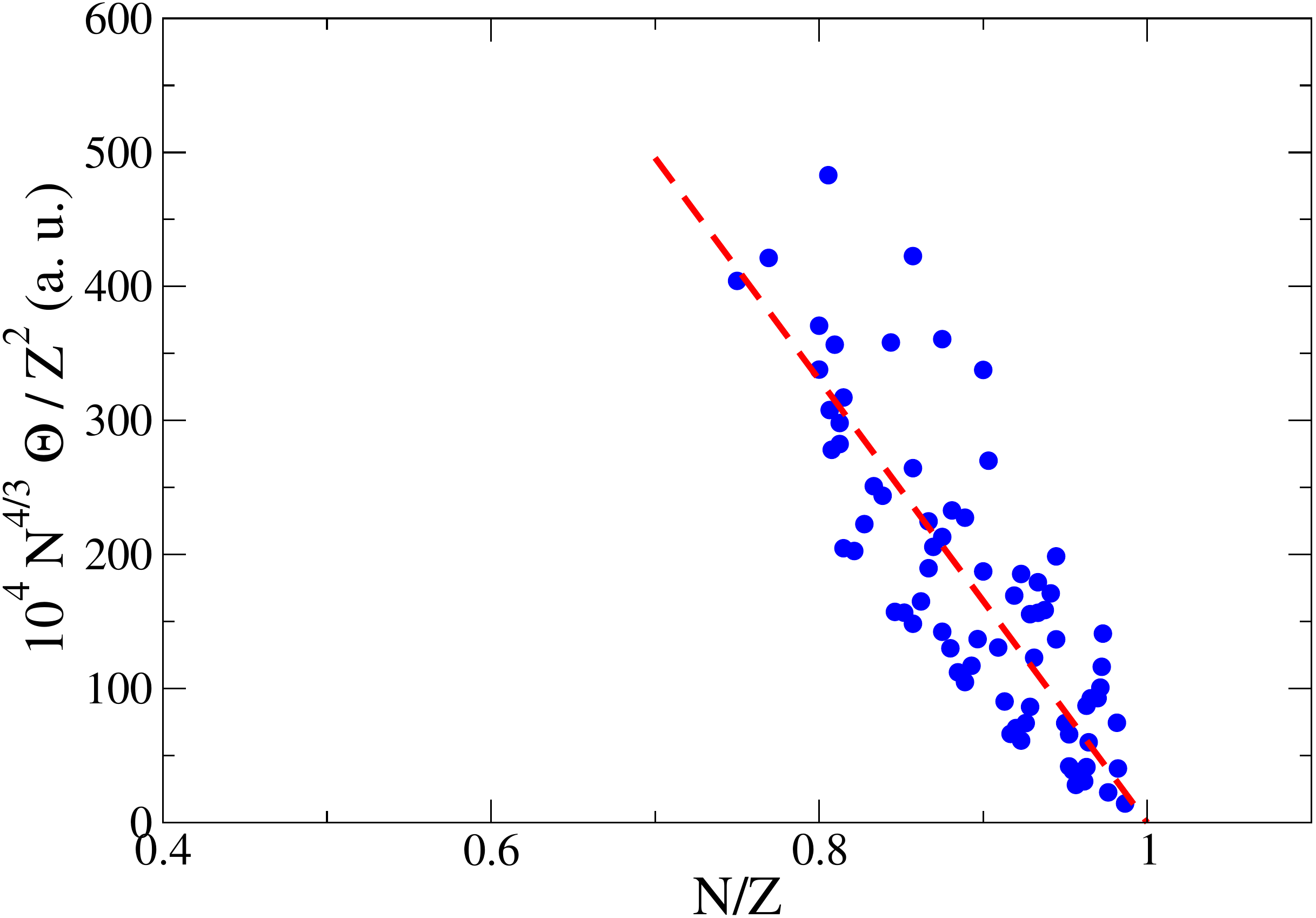}
\caption{\label{fig3} (Color online) Scaling of the temperature
parameter, $\Theta$. Ions with $N>10$ and at least 100 measured lines 
are included in the plot.}
\end{center}
\end{figure}

First, we notice that from Ar I to Ca III the spectrum is restructured.
This corresponds to reestablishing the $Z\to\infty$ sequence of
Hartree-Fock orbitals: $1s$, $2s$, $2p$, $3s$, $3p$, $3d$,\dots, in 
place of the neutral atom sequence \cite{MC}.

From Ca III on, it is evident that the low-lying spectrum goes to a 
definite limit as $Z$ increases. These are intra-shell excitations 
(in the $N=18$ system, the $n=3$ shell is open). This 
means that intra-shell excitation energies scale as $\Delta E\sim Z^2$,
independently of the degree of ionization, $N/Z$. We notice that from 
$Z\to\infty$ perturbation theory one would expect intra-shell
excitations to be of the order of $Z$, thus the observed behaviour is
stressing that the lowest excitations have a non-perturbative
character.

Larger excitation energies correspond to inter-shell transitions, which
are of the same order of $E_{ioniz}$ and, thus, exhibit the
characteristic dependence on $Z$, $N$, and $N/Z$. 

Let us stress that
for both the intra- and the inter-shell regimes, locally the 
number of states increases exponentially with energy differences. This
behaviour is quite general \cite{universality}. It was first observed in
nuclear systems, where it is known as the ``constant temperature 
approximation'' \cite{CTA}. In Fig. \ref{fig2}, dashed lines represent
fits to the intra- and inter-shell excitations in the Mn VIII ion. The
discontinuity of the slopes signals that the mechanisms of formation of
these states are different\cite{Capote}.

A rough characterization of the spectrum can be given in terms of a
single exponential:

\begin{equation}
N_{states}\approx \exp (\Delta E/\Theta), 
\label{eq4}
\end{equation}

\noindent
where we expect the temperature parameter, $\Theta\sim Z^2$. The
dependence of $\Theta$ on $N$ can be extracted from the $Z\to\infty$
asymptote. In this limit and for closed shell ions, the gap and the
number of electrons are expressed in terms of the principal quantum
number of the last occupied shell as:  $\Delta E_1\approx Z^2/n^3$, and
$N\approx 2 n^3/3$, respectively. The available number of levels at
energies $\Delta E_1$ is $N_{states}\sim [2(n+1)^2]!/\{[2 n^2]!~
[2(2 n+1)]!\}\sim\exp(a n+b n\ln n)\sim \exp(c n)$, where $a$, $b$, and 
$c$ are numerical coefficients. The last expression comes from the fact
that $\ln n$  is a slowly varying function. Comparison with Eq.
(\ref{eq4}) leads to $\Theta\sim Z^2 N^{-4/3}$.

In the spirit of Thomas-Fermi scaling, we assume for $\Theta$ the 
relation:

\begin{equation}
\Theta=Z^2 N^{-4/3} h(N/Z), 
\end{equation}

\noindent
where $h$ is a universal function. We fit the excitation spectra of ions
with $N>10$ with a function given by Eq. (\ref{eq4}). We study all of
the ions with
at least 100 measured lines, and use the first 100 lines to perform the 
fit. The results are shown in Fig. \ref{fig3}.

Although Eq. (\ref{eq4}) gives a rough global characterization (the 
exponential dependence holds locally, with different slopes in different
sectors of the spectrum), and there could be experimental problems such 
as missing lines, etc, the data shows scaling, with a function $h$ 
approximately linear:

\begin{equation}
h\approx 0.165~(1-N/Z).
\end{equation}

\noindent
Due to the reasons mentioned above, the observed dispersion of points 
is natural.

Similarly to the ionization potential, the parameter $\Theta$ for
neutral atoms takes in mean a constant value, $\Theta\approx 0.045$,
with strong fluctuations related to shell effects. The reason for such a
behaviour is also the proximity of the anionic instability border.

In conclusion, we verified that Thomas-Fermi theory offers a fresh view
to the energy spectra of atomic ions. The energy spectrum can be
described in a ``universal'' way in terms of scaling functions. One of
these functions describes the first ionization potential, which is the 
natural scale for the excitation spectrum. On the other hand, if the
excitation spectrum is fitted by a single exponential, the temperature
parameter is shown to approximately exhibit also a scaling relation 
{\it a la} Thomas-Fermi. 

For neutral atoms, the scaling is broken because of the proximity of the
anionic instability border. We shall further study this border seeking
for positive uses of its presence.

There are still many interesting points requiring further investigation.
Intra- and inter-shell excitations could be more detailed
described by means of constant-temperature approximations. Alternative
characterizations of the spectrum could also be employed. We have in
mind, for example, the first moment, which is a magnitude with
dimensions of energy. The energy region closer to the ionization
potential could also be studied. In this region, we expect the
dependence $N_{states}\sim (E_{ioniz}-\Delta E)^{-3/2}$, coming from the
existence of Rydberg states. A very interesting possibility is to write
down an interpolation formula to spectroscopic accuracy for the 
ionization potential, which takes our $g$ function as the starting 
point, but includes also shell-filling effects, spin-orbit interactions,
etc. Research along these lines is in progress.

\begin{acknowledgments}
The authors are grateful to the participants of the Theory Seminar at
the ICIMAF for discussions and remarks. Support by the Caribbean 
Network for Quantum Mechanics, Particles and Fields (ICTP, Trieste, 
Italy), and by the Programa Nacional de Ciencias B\'asicas (Cuba) is
acknowledged.
\end{acknowledgments}

\end{document}